\documentclass[preprint,showpacs,preprintnumbers,amsmath,amssymb,superscriptaddress]{revtex4}

\usepackage{graphicx}
\usepackage{dcolumn}
\usepackage{bm}
\begin{document}

\title{Two-body Correlations and the Superfluid Fraction for Nonuniform Systems}
\author{W.M. Saslow}
\affiliation{ Department of Physics, Texas A\&M University, College Station, TX 77843-4242}
\author{D.E. Galli}
\affiliation{ Dipartimento di Fisica, Universit\`a degli Studi di Milano, Via Celoria 16, 20133 Milano, Italy}
\author{L. Reatto}
\affiliation{ Dipartimento di Fisica, Universit\`a degli Studi di Milano, Via Celoria 16, 20133 Milano, Italy}
\date{\today}


\begin{abstract}
We extend the one-body phase function upper bound on the superfluid fraction $f_s$ in a periodic solid (a spatially ordered supersolid) to include two-body phase correlations.  
The one--body current density is no 
longer proportional to the gradient of the one-body phase times
the one--body density, but 
rather it becomes
$\vec{j}(\vec{r}_1)=\rho_{1}(\vec{r}_1)\frac{\hbar}{m}\vec{\nabla}_{1}\phi_{1}(\vec{r}_1)
+\frac{1}{N}\int d\vec{r}_2\rho_{2}(\vec{r}_1,\vec{r}_2)\frac{\hbar}{m}\vec{\nabla}_{1}
\phi_{2}(\vec{r}_1,\vec{r}_2)$.
This expression therefore depends also on two-body correlation functions.  
The equations that simultaneously determine the one-body and two-body phase functions require a knowledge of one-, two-, and three-body correlation functions.  The approach can also be extended to disordered solids.  Fluids, with two-body densities and two-body phase functions that are translationally invariant, cannot take advantage of this additional degree of freedom to lower their energy.  
\end{abstract}


\maketitle

\section{INTRODUCTION}
In 1970, Leggett suggested that solid $^4$He might exhibit non-classical rotational 
inertia (NCRI).\cite{Leggett1}  Recently, Kim and Chan have observed this effect for 
solid $^4$He in Vycor, porous gold, and in bulk.\cite{KCvycor,KCgold,KCbulk}  Given 
that, despite the extremely different porosities of these samples, the extrapolated 
values of the temperature $T=0$ K NCRI fraction (or NCRIF) are all in the 1-2\% 
range, it seems unlikely that the effect occurs only at surfaces.  Other experiments 
reproduce these results qualitatively, but with slightly smaller values, according to 
how the sample is prepared.\cite{Reppy,Shira,Kubo}  
In all previous experiments annealing has been used to improve the quality of the solid samples, but only
in one case is it reported that the effect can disappear under annealing.\cite{Reppy}

We have recently computed, using one-body densities obtained with an exact $T$=0 K Quantum Monte Carlo method, the superfluid fraction upper bound due to Leggett, based on a one-body phase function and the one-body density\cite{GalliRS}.  Our values were of the order of 20\%, far above what has been observed.  If supersolidity indeed does occur, there is a need to understand why it does so at much lower values.  For that reason, and because 
superfluidity in presence of a traslational broken symmetry may occur under other circumstances (such as condensates in optical lattices), we have re-examined the question of superfluidity in a supersolid, permitting the system to develop a phase function that includes not merely one-body effects, but also two-body (and, in principle, higher) terms.

Leggett's original study considered superflow in a solid confined to an annulus. 
He first ruled out systems whose ground state wavefunction $\Psi_{0}$ is ``disconnected'' 
when a single particle coordinate is moved around the annulus (a generalized insulating state), and 
systems for which the wavefunction re-organizes to a lower energy state (a generalized 
form of the Landau criterion).  To obtain an upper limit on the NCRIF, he then explicitly 
considered that, under rotation, the wavefunction became $\Psi=\Psi_{0}\exp({i\phi})$, with a phase function $\phi$ that was a sum of terms depending only upon a single 
coordinate (what we call a one-body phase function).\cite{Leggett1}  He noted that this would yield only an upper limit to the $T=0$ K superfluid fraction $f_{s}$.  Slightly prior to 
Leggett, Andreev and Lifshitz posited superfluidity due to quantum diffusion of unspecified defect excitations.\cite{A&L}  A macroscopic theory of superflow was developed, but no microscopic definition of the superfluid velocity was given.  At about the same time 
Chester\cite{chester} noted the existence of variational ground state wave functions which
display translational broken symmetry, corresponding to a crystal with vacancies;  
for such wave functions Bose Einstein condensation (BEC) had been demonstrated\cite{reatto}.
Ref.~\onlinecite{A&L} suggested that capillary flow with a fixed lattice might occur under gravity, and one might expect other possibilities for unusual flow associated with solid $^4$He.  Despite numerous efforts, no such flow has been observed.\cite{Andreev,Greywall,Beamish,Meisel}  Additional, primarily topological, considerations have been given by Leggett,\cite{Leggett2} who in more recent work  
introduces the symbol $f_s$ for the superfluid fraction (SFF).\cite{Leggett3}  If superflow is the 
correct interpretation of the NCRI experiments, then $f_{s}$ is equivalent to the NCRIF. 

At the microscopic level, a number of Monte Carlo calculations for solid $^4$He have been 
performed, with no common conclusion.  Ceperley and Bernu have performed
calculations on solid $^4$He at low finite temperature 
with path integral Monte Carlo computations
and to date they find 
evidence that is contrary to superfluidity.\cite{CeperleyBernu}  They employ 
the winding number method, which yields reasonable values for the 
superfluid properties of liquid $^4$He.  For smaller samples they find a finite $f_{s}$, 
but for larger samples they find $f_{s}=0$, although they observe that this negative result
from the winding number calculation for the larger samples is 
inconclusive, due to possible phase-space sampling problems.  They also calculate the exchange integrals, which their method successfully obtains for solid $^3$He, and they interpret their 
results as arguing against superfluidity.  Prokofiev and Svistunov also argue 
against bulk superfluidity.\cite{ProkoSvis1}  On the other hand, Galli, Rossi, 
and Reatto 
studied off-diagonal long-range order (ODLRO) at zero temperature
for relatively large systems on the basis of an accurate variational theory, 
finding small but definite asymptotic values, which imply a finite BEC in the zero-momentum 
state, and ``connectivity'' of the wavefunction.\cite{GalliRR}  As a consequence, 
a superfluid velocity can be defined, and one can expect superflow if the excitations satisfy the Landau criterion.  More recently, it has been found that superfluidity can occur in Monte Carlo calculations for a disordered ``sample''.\cite{Boninsegni}  Such a sample was not equilibrated, and was deliberately disordered, and thus is far from equilibrium.  One cannot rule out the possibility that this occurs experimentally, although it is not clear that the experimental samples are as disordered as the simulated samples. 
An exact computation\cite{GalliPRL} 
at zero temperature for a crystal with a finite 
concentration of vacancies gave ODLRO and therefore BEC but again it is not
known if the experimental samples contain a finite concentration of vacancies at the
low temperature of the experiment.

Having ODLRO and satisfying the Landau criterion should ensure superfluidity, 
but it still leaves open the problem of computing the value of $f_s$; this amounts
to finding the best phase function, 
in the sense of minimizing the energy associated with superflow, where the phase 
gradient gives the superfluid velocity.  The best such one-body phase function 
(but perhaps not the best phase function) satisfies the continuity equation 
with the current given by the product of the one-body density and the 
one-body superfluid velocity.\cite{Saslow1}  This provides an equation that 
enables the superfuid velocity profile to be determined from the average 
superfluid velocity and the local one-body density profile.  Using a model 
for hcp $^4$He as an fcc lattice of the same density, ref.~\onlinecite{Saslow1} 
took the one-body density to be a sum of gaussians.  For relatively localized 
gaussians, the original calculations were not converged, due to computer 
limitations at that time, and thus provided only an upper limit for the 
superfluid fraction from the optimized one-body function.  
Using a 1976 estimate of the best gaussian, and more extensive computation, 
ref.~\onlinecite{Saslow2} gave an upper bound for $f_s$, or NCRIF, 
on the order of 2\%, apparently in good agreement with experiment.  However, more 
accurate Gaussian approximations to the one-body density, determined from 
"exact" Monte Carlo 
one-body densities, and the one-body densities themselves, have recently been shown to give $f_s$ on the order of 20\% for solid $^4$He at density of 0.029 \AA$^{-3}$ (near the melting pressure).\cite{GalliRS}  This is clearly too large a value, and requires re-examination of the theory.  In what follows we therefore consider the possibility that a superfluid can respond to rotation by developing a phase that also includes two-body (and, in principle, higher) terms.  In this case the superflow for each particle reflects not merely the local number density $\rho(\vec{r})$, but also 
correlations with the other particles.  

Thus, as a particle flows around the crystal, it sees not only the average mass 
density, but also details of how other particles correlate to it, and changes its 
phase in response to them.  When the two-body phase term is included, the 
one-body current density, which specifies the superflow pattern, is more complex than the usual $\vec{j}(\vec{r})=\rho(\vec{r})\vec{v}_{s}(\vec{r})$ form; it now depends explicitly on the two-body density (see Eq.~\ref{newonebodycurrent}), and implicitly on both the two- and three-body 
densities.  Although the former is occasionally computed (for example, in 
correlated basis-function studies), the three-body density is not normally 
studied.  In the absence of such information we merely present the 
formalism, and do not perform actual calculations.
Neverthess, this extra degree of freedom 
should enable the solid to lower its flow energy, and therefore to decrease the $f_s$.   
This extension of the phase function to include two-body and higher terms should also 
apply to other candidates for supersolid behavior, such as bosons confined in optical lattices. 

For bulk superfluid, the translational invariance of certain integrals involving 
the two-body correlation function and the two-body phase function prevents the system from utilizing the 
two-body phase term.  This explains why, for a bulk superfluid at $T=0$, 
theory based on a one-body phase alone gives the maximum possible value for $f_{s}$, which is unity.  

Note that, if superfluidity is a bulk phenomenon in disordered Bose systems, 
then with certain developments the method of the present work should give the two-body upper bound for $f_s$.  The advantage of having a means to compute 
also upper bounds on $f_s$, as opposed to only Monte Carlo calculations of $f_s$, 
is that it ensures consistency and reveals the extent to which correlations are involved in the superfluidity. 

\section{ENERGY OF SUPERFLOW FOR A NON-UNIFORM SYSTEM WITH $N$-BODY CORRELATIONS}

In what follows $x_i$ will be used to represent the vector position $\vec{x}_i$ of the $i$-th particle.  Consider $N$ identical Bose particles of mass $m$, with ground-state wavefunction $\Psi_{0}(x_{1}\dots x_{N})$ and energy $E_{0}$.  $\Psi_{0}$ is an eigenstate of the Hamiltonian  
\begin{equation}
H=-\frac{\hbar^2}{2m}\sum_{i}^{N}\nabla_{i}^{2}+V, 
\label{H}
\end{equation}
where $V$ is a sum of two-body interactions.   
Because $\Psi_{0}$ is the ground state, it can be considered to be a real function, 
up to a constant phase.  Our goal is to find the energy $E$ of the system when there 
is an imposed average superflow velocity $v_{0}$.  The energy $E-E_{0}$ provides a 
measure of the superfluid fraction. 

Specifically, we consider superflow for a periodic geometry with average velocity $\vec{v}_0$, 
with macroscopic repeat distance $L$ and cross-section $A$.  This imposes a phase 
change of $mv_{0}L/\hbar$ on taking a single particle through $L$.  Following Leggett, we assume that the effect on the wavefunction is simply to change its phase, via 
\begin{equation}
\Psi_{0}\rightarrow\Psi=\Psi_{0}\exp(i\phi),
\label{rotbc}
\end{equation}
where $\phi$ will depend on all 
$N$ coordinates, perhaps in a rather complex fashion.  Leggett has also considered 
additional possibilities that we do not discuss here.\cite{Leggett2,Leggett3}  
Note that for such a response
neither the density nor the potential energy changes; 
only the kinetic energy changes.  

With normalization $\int|\Psi_{0}|^{2}dx_{1}\dots dx_{N}=1$, it is straightforward to show that 
\begin{equation}
\Delta E=E-E_{0}
=\frac{\hbar^2}{2m}\sum_{i}^{N}\int|\Psi_{0}|^{2}(\vec{\nabla}_{i}\phi)^{2}dx_{1}\dots dx_{N}.
\label{Schrod4}
\end{equation}
Minimization of $\Delta E$ with respect to a $\phi$ that depends on all the particle 
coordinates ($N$-body correlations) leads to 
$\vec{\nabla}_{i}\cdot(|\Psi_{0}|^{2}\vec{\nabla}_{i}\phi)=0$ for each $i$.

\section{SUPERFLOW FOR A NON-UNIFORM SYSTEM WITH TWO-BODY CORRELATIONS}

Now consider a $\phi$ that depends on correlations between the coordinates.  Let us expand it in the form 
\begin{equation}
\phi=\sum_{i}^{N}\phi_{1}(x_{i})+\frac{1}{2(N-1)}\sum_{i\ne j}^N\phi_{2}(x_{i},x_{j})+\dots
\label{phi-expand}
\end{equation}
The one-body term includes both the imposed flow and the one-body backflow.  The two-body contributions to the phase represent correlated backflow, and dots represent more complex backflow effects.
By symmetry, $\phi_{2}(x_{i},x_{j})=\phi_{2}(x_{j},x_{i})$.  There is no constant term in 
either coordinate, since that would be part of $\phi_{1}$.  Hence, in the fourier expansion 
of $\phi_{2}$, there are no zero-wavevector terms.  Many-body phases have been considered 
previously, but not in the context of steady bulk superflow.\cite{KroTym,VitiReattoKalos,OrtizCeperley} 

In terms of $\phi$, which must be symmetric in all coordinates, we have 
\begin{equation}
\vec{v}_{s1}(x_{1},\dots)\equiv\frac{\hbar}{m}\vec{\nabla}_{1}\phi
=\vec{v}_{s}(1)+\frac{1}{N-1}\sum_{i\ne1}^N\vec{v}_{s}(1,i)+\dots,
\label{superv1}
\end{equation}
where $\vec{v}_{s}(1)\equiv\frac{\hbar}{m}\vec{\nabla}_{1}\phi_{1}(x_{1})$ and 
$\vec{v}_{s}(1,i)\equiv\frac{\hbar}{m}\vec{\nabla}_{1}\phi_{2}(x_{1},x_{i})$ 
are expected to be of the same order of magnitude.  
If the $N-1$ two-body velocities are to yield a net velocity that is comparable 
to those from the one-body velocity, then the two-body velocity should be on 
the order of the one-body velocity. 

We then have 
\begin{eqnarray}
\vec{v}^{\,2}_{s1}=\vec{v}_{s}(1)^2&+&\frac{2}{N-1}\vec{v}_{s}(1)\cdot\sum_{i\ne1}^N\vec{v}_{s}(1,i)
+(\frac{1}{N-1})^2\sum_{i\ne1}^N\vec{v}^2_{s}(1,i) \nonumber \\
&+&(\frac{1}{N-1})^2\sum_{i\ne1}^N\vec{v}_{s}(1,i)\cdot\sum_{j\ne1,i}^N\vec{v}_{s}(1,j)+\dots.
\label{superv2}
\end{eqnarray}
The first term is what one would obtain from a one-body phase.  The second is a (new) cross-term, 
involving both the one-body and two-body phases, that makes the one- and two-body phases dependent on one another.  
(There are $N-1$ equal terms of this sort.)  The third is the square of the two-body term 
involving 1 and $i$.  (There are $N-1$ equal terms of this sort; in the limit  
$N\rightarrow\infty$, this term is negligible so that it is dropped from now on.)  
The fourth term gives the two-body phase cross-correlations.  
(There are $2(N-1)(N-2)$ equal terms of this sort, so it dominates the contribution of the 
third term.)  

It is now convenient to define a number of quantitites.  The one-body density $\rho_{1}$ is given by
\begin{equation}
\rho_{1}(1)\equiv\rho_{1}(x_{1})\equiv N\int|\Psi_{0}|^{2}dx_{2}\dots dx_{N},
\label{one-body}
\end{equation}
where $\int dx_{1}\rho_{1}(x_{1})=N$.  The two-body density $\rho_{2}$ is given by 
\begin{equation}
\rho_{2}(1,2)\equiv\rho_{2}(x_{1},x_{2})\equiv N(N-1)\int|\Psi_{0}|^{2}dx_{3}\dots dx_{N},
\label{two-body}
\end{equation}
where $\int \rho_{2}(x_{1},x_{2})dx_{2}=(N-1)\rho_{1}(x_{1})$.  The three-body density $\rho_{3}$ is given by  
\begin{equation}
\rho_{3}(1,2,3)\equiv\rho_{3}(x_{1},x_{2},x_{3})\equiv N(N-1)(N-2)\int|\Psi_{0}|^{2}dx_{4}\dots dx_{N}, 
\label{three-body}
\end{equation}
where $\int \rho_{3}(x_{1},x_{2},x_{3})dx_{3}=(N-2)\rho_{2}(x_{1},x_{2})$.  With 
these conventional definitions, one has $\rho_{N}(1,\dots,N)=N!|\Psi_{0}|^{2}$, which 
leads to additional factors of order $N$ that later appear in the theory of the flow energy.  
For $n\ll N$, note that $\rho_n\sim (N/V)^n$.  

In terms of these densities, 
and in the limit $N\rightarrow\infty$, if the one-body and two-body velocities are of the same order 
of magnitude, we obtain for the flow energy
\begin{eqnarray}
\Delta E\rightarrow\frac{m}{2}\int\rho_{1}(x_{1})(\vec{v}_{s}(1))^2dx_{1}
+\frac{m}{N}\int\rho_{2}(x_{1},x_{2})\vec{v}_{s}(1)\cdot\vec{v}_{s,2}dx_{1}dx_{2}\nonumber \\
+\frac{m}{2N^{2}}\int\rho_{3}(x_{1},x_{2},x_{3})\vec{v}_{s,2}\cdot\vec{v}_{s,3}dx_{1}dx_{2}dx_{3}.
\label{Schrod6N}
\end{eqnarray}
The second and the third terms of this equation represent correlated backfow contributions 
the flow energy.  In the many-body expansion of $\phi$ we have truncated correlated backflow to the pair level; these results would be more complex if triplet correlations were included in $\phi$. 

\section{MINIMIZATION OF THE ENERGY OF SUPERFLOW}

The condition that $E$ be a minimum relative to variations in $\phi_{1}(x_{1})$ leads to 
\begin{eqnarray}
0=\frac{1}{\hbar}\frac{\delta\Delta E}{\delta\phi_1}
&=&\vec{\nabla}_{1}\cdot[\rho_{1}(x_{1})\vec{v}_{s}(1)+\frac{1}{N}\int dx_{2}\rho_{2}(x_{1},x_{2})\vec{v}_{s}(1,2)]\nonumber \\
&=&\vec{\nabla}_{1}\cdot\vec{j}(x_{1}), 
\label{continuityN1}
\end{eqnarray}
where the one body current density $\vec{j}(x_{1})$ now is the sum of the usual one-body term and an integral over the two-body density:  
\begin{equation}
\vec{j}(x_{1})=\rho_{1}(x_{1})\vec{v}_{s}(1)+\frac{1}{N}\int dx_{2}\rho_{2}(x_{1},x_{2})\vec{v}_{s}(1,2).
\label{newonebodycurrent}
\end{equation} 

The condition that $E$ be a minimum relative to variations in $\phi_{2}(x_{1},x_{2})$ 
leads to (in the limit $N\rightarrow\infty$)
\begin{equation}
0=\frac{1}{\hbar}\frac{\delta\Delta E}{\delta\phi_2}
=\vec{\nabla}_{1}\cdot[\rho_{2}(x_{1},x_{2})\vec{v}_{s}(1)
+\frac{1}{N}\int dx_{3}\rho_{3}(x_{1},x_{2},x_{3})\vec{v}_{s}(1,3)].
\label{continuityN2}
\end{equation}

With $V$ the system volume, 
%
and with periodic boundary conditions
we expand the known densities via 
\begin{eqnarray}
\rho_{1}(x_{1})&=&\sum_{\lambda}\rho_{\lambda}\exp[i\vec{G}_{\lambda}\cdot\vec{x}_{1}], \label{expandrhoa}\\
\rho_{2}(x_{1},x_{2})&=&\frac{N}{V}\sum_{\lambda\mu}\rho_{\lambda,\mu}\exp[i(\vec{G}_{\lambda}\cdot\vec{x}_{1}+\vec{G}_{\mu}\cdot\vec{x}_{2})],\label{expandrhob}\\
\rho_{3}(x_{1},x_{2},x_{3})
&=&\frac{N^{2}}{V^{2}}\sum_{\lambda\mu\nu}\rho_{\lambda,\mu,\nu}\exp[i(\vec{G}_{\lambda}\cdot\vec{x}_{1}+\vec{G}_{\mu}\cdot\vec{x}_{2}+\vec{G}_{\nu}\cdot\vec{x}_{3})].
\label{expandrho}
\end{eqnarray}
If the system is in a crystalline phase and the center of mass is fixed, then the densities must reflect
the periodicity of the crystal lattice. In this case the vectors $\vec{G}_{\lambda}$,
$\vec{G}_{\lambda}+\vec{G}_{\mu}$, and $\vec{G}_{\lambda}+\vec{G}_{\mu}+\vec{G}_{\nu}$
must be vectors of the reciprocal lattice.
These definitions give all the fourier coefficients (e.g. $\rho_{\lambda}, \rho_{\lambda,\mu}$) 
the same dimensionality.  Note that $\rho_0=N/V$ is the average particle density.  

We now expand the unknown phases via 
\begin{eqnarray}
\phi_{1}(x_{1})&=&\frac{m\vec{v}_{0}\cdot\vec{x}_{1}}{\hbar}+\sum_{\lambda\ne0}\phi_{\lambda}\exp[i\vec{G}_{\lambda}\cdot\vec{x}_{1}],
\label{expandphi1}\\
\phi_{2}(x_{1},x_{2})&=&\sum_{\lambda\ne0,\mu\ne0}\phi_{\lambda,\mu}\exp[i(\vec{G}_{\lambda}\cdot\vec{x}_{1}+\vec{G}_{\mu}\cdot\vec{x}_{2})].
\label{expandphi2}
\end{eqnarray}
Alternatively, in (\ref{expandphi1}) we can sum over all $\lambda$ by taking $\vec{G}_{0}$ to be 
very small and $\phi_{0}$ to be very large, subject to the condition that 
$\vec{v}_{s0}\equiv({i\hbar/m})\vec{G}_{0}\phi_{0}$ be finite.  The first term in (\ref{expandphi1}) is the imposed flow, and the second term is the (to-be-determined) one-body backflow. 

In practice, we consider $\vec{v}_{0}$ to be known, and we must find the other components 
in terms of $\vec{v}_{0}$ or, equivalently, $\phi_{0}$.  The unknown velocities become, 
via $\vec{v}_{s}=(\hbar/m)\vec{\nabla}\phi$,  
\begin{eqnarray}
\vec{v}_{s}(x_{1})&=&\vec{v}_{0}+\frac{i\hbar}{m}\sum_{\lambda\ne0}\vec{G}_{\lambda}\phi_{\lambda}\exp[i\vec{G}_{\lambda}\cdot\vec{x}_{1}]\nonumber \\
&=&\vec{v}_{0}+\sum_{\lambda\ne0}\vec{v}_{\lambda}\exp[i\vec{G}_{\lambda}\cdot\vec{x}_{1}],\\
\vec{v}_{s}(x_{1},x_{2})&=&\frac{i\hbar}{m}\sum_{\lambda\mu}\vec{G}_{\lambda}\phi_{\lambda,\mu}\exp[i(\vec{G}_{\lambda}\cdot\vec{x}_{1}+\vec{G}_{\mu}\cdot\vec{x}_{2})]\nonumber \\
&=&\sum_{\lambda\ne0,\mu\ne0}\vec{v}_{\lambda,\mu}\exp[i(\vec{G}_{\lambda}\cdot\vec{x}_{1}+\vec{G}_{\mu}\cdot\vec{x}_{2})], 
\label{expandv}
\end{eqnarray}
where $\vec{v}_{\lambda}\equiv(i\hbar/m)\vec{G}_{\lambda}\phi_{\lambda}$, and 
$\vec{v}_{\lambda,\mu}\equiv(i\hbar/m)\vec{G}_{\lambda}\phi_{\lambda,\mu}$.  

Eq.~(\ref{continuityN1}) implies that, for all $\lambda\ne0$, that  
\begin{equation}
0=\sum_{\mu}(\vec{G}_{\lambda}\cdot\vec{G}_{\mu})
[\rho_{\lambda-\mu}\phi_{\mu}+\sum_{\gamma\ne0}\rho_{\lambda-\mu,\gamma}\phi_{\mu,-\gamma}].
\label{continuityN1G}
\end{equation}
For a discretization where $\lambda$ takes on $P$ values, this gives $P-1$ conditions on $P-1$ 
unknown values of $\phi_{\lambda}$ (where $\phi_{0}$ is considered to be known).  
For $\phi_{\mu,-\gamma}=0$, these are the equations employed in refs.~\onlinecite{Saslow1,Saslow2,Saslow-hcp}.

Eq.~(\ref{continuityN2}) implies that, for all $\lambda\ne0$ and all $\nu\ne0$, that 
\begin{equation}
0=\sum_{\mu}(\vec{G}_{\lambda}\cdot\vec{G}_{\mu})
[\rho_{\lambda-\mu,\nu}\phi_{\mu}+\sum_{\gamma\ne0}\rho_{\lambda-\mu,\nu,\gamma}\phi_{\mu,-\gamma}].
\label{continuityN2G}
\end{equation}
Note that all of the $\phi$'s are driven by the superflow, and thus must be proportional 
to $\vec{v}_{0}$ dotted with a vector associated with the system.  For a discretization 
where $\lambda$ takes on $P$ values and $\nu$ takes on $P$ values, this gives $(P-1)^2$ conditions 
on $(P-1)^2$ unknown values of $\phi_{\mu,-\gamma}$.  

\section{SUPERFLUID DENSITY}

We now derive an expression for the superfluid density.  (Recall that this is really an upper limit to the superfluid density.) 
Let us rewrite (\ref{Schrod6N}), for $N\rightarrow\infty$, as
\begin{eqnarray}
\Delta E
=\frac{m}{2}\int dx_{1}\vec{v}_{s}(1)\cdot[\rho_{1}(x_{1})\vec{v}_{s}(1)
+\frac{1}{N}\int dx_{2}\rho_{2}(x_{1},x_{2})\vec{v}_{s}(1,2)]\nonumber\\
+\frac{m}{2N}\int dx_{1}dx_{2}\vec{v}_{s}(1,2)\cdot\vec{v}_{s}(1)\rho_{2}(x_{1},x_{2})\nonumber\\
+\frac{m}{2N^2}\int dx_{1}dx_{2}dx_{3}\vec{v}_{s}(1,2)\cdot\vec{v}_{s}(1,3)\rho_{3}(x_{1},x_{2},x_{3}).
\label{Schrod7N}
\end{eqnarray}
Because $\vec{v}_{s}(1)$ and $\vec{v}_{s}(1,2)$ involve $\vec{\nabla}_{1}$, in both of these 
terms we can integrate by parts, where we take the integration direction along the flow.  
Each volume term then leads to a surface term and to a volume term that is the product of a phase with a divergence.  
However, by (\ref{continuityN1}) and (\ref{continuityN2}), these divergences are zero.  
Moreover, the quantity $\vec{v}_{s}(1,2)$ is strictly periodic, so its phase is periodic, 
and therefore its two surface terms cancel.  All that remains is the surface term 
associated with the change in phase of $\vec{v}_{s}(1)$ as a particle goes around the system.  
Defining $\Sigma_1$ to be this surface, of area $A$, and defining $\hat{v}_0$ to be the unit vector normal to $\Sigma_1$,
i.e. the direction of the imposed velocity $\vec{v}_0$,
we then have 
\begin{eqnarray}
\Delta E&=&\frac{\hbar}{2}\int_{\Sigma_1} d{A}_{1}(\Delta\phi_{0})\hat{v}_{0}\cdot[\rho_{1}(x_{1})\vec{v}_{s}(1)
+\frac{1}{N}\int dx_{2}\rho_{2}(x_{1},x_{2})\vec{v}_{s}(1,2)]\nonumber\\
&=&\frac{\hbar}{2}A(\Delta\phi_{0})j_{0}=\frac{m}{2}ALv_{0}j_{0}\equiv\frac{m}{2}V\rho_{s}v_{0}^{2},
\label{Schrod8N}
\end{eqnarray}
where $V=AL$ and $j_{0}=\rho_{s}v_{0}$ is given in fourier component form as 
\begin{eqnarray}
\vec{j}_{0}&=&\frac{i\hbar}{m}\sum_{\mu}\vec{G}_{\mu}[\rho_{-\mu}\phi_{\mu}+\sum_{l\ne0}\rho_{-\mu,l}\phi_{\mu,-l}]\nonumber \\
&=&\rho_{0}\vec{v}_{0}+\frac{i\hbar}{m}{\sum_{\mu\ne0}}\vec{G}_{\mu}[\rho_{-\mu}\phi_{\mu}+\sum_{l\ne0}\rho_{-\mu,l}\phi_{\mu,-l}].
\label{j_0}
\end{eqnarray}
There is no contribution to $j_{0}$ from the part of $\rho_{2}(x_{1},x_{2})$ that is 
independent of $x_{1}$ because the integral over $x_{1}$ involves a strictly periodic 
function of $x_1$.  Hence, once we know the flow pattern in terms of $v_{0}$, we 
can compute $j_{0}$ and determine $\rho_{s}=j_{0}/v_{0}$.  The first term in (\ref{j_0}) is the imposed flow, the second term is the one-body backflow, and the third term is the two-body backflow. 

The theory is sufficently general that it can be applied both to liquid and solid boson systems.
For a homogeneous liquid the previous equations simplify drastically.  In the liquid 
the two-body density $\rho_{2}(x_{1},x_{2})$ is a function only 
of $|\vec{x}_{1}-\vec{x}_{2}|$.  For the liquid we expect the two-body phase $\phi_{2}(x_{1},x_{2})$ 
also to be a function only of $|\vec{x}_{1}-\vec{x}_{2}|$.  As a consequence, 
$\rho_{2}(x_{1},x_{2})$ is even in $\vec{x}_{1}-\vec{x}_{2}$, but $\vec{v}_{s}(1,2)$, 
involving $\vec{\nabla}_{1}|\vec{x}_{1}-\vec{x}_{2}|$, is odd in $\vec{x}_{1}-\vec{x}_{2}$.  
Hence the second term in (\ref{Schrod8N}) goes to zero.  Therefore the two body terms 
do not enter explicitly in determining $\rho_{s}$.  Moreover, in (\ref{continuityN1}), 
the two-body terms similarly are zero, so that they do not even affect the one-body terms.  
Hence the two-body terms, i.e. backflow, are irrelevant both to determining the one-body flow pattern and 
the total energy of a uniformly flowing translationally invariant system.  
Therefore it is valid to neglect two-body terms in determining the superfluid 
density in a fluid, as is done conventionally.  We believe that study of the 
higher-order phase correlations will reveal that none of them are relevant to a superfluid.  The one-body phase alone then gives a 
$T=0$ K superfluid fraction of unity, in agreement with a conclusion in ref.~\onlinecite{Leggett3}. 

The actual computation of the upper bound on $f_{s}$ which includes the two-particle phase requires the solution of eq.(29), where the phases $\phi_{\mu}$ and $\phi_{\mu,-\gamma}$ are solutions of eqs.(25) and (26).  As inputs one needs to know $\rho_{\lambda}, \rho_{\lambda,\mu}$, and $\rho_{\lambda,\mu,\nu}$, the fourier components of the one-body, two-body, and three--body correlation functions, respectively, as defined in eqs.(18-20).  
There is no difficulty in obtaining highly accurate values for $\rho_{\lambda}$ from an exact
computation\cite{GalliRS} for the ground state of solid $^4$He.  Also, the computation of $\rho_{\lambda,\mu}=\frac{1}{N}<\rho_{\vec{G}_{\lambda}}\rho_{\vec{G}_{\mu}}>$ is currently performed in Monte Carlo simulations and there is no difficulty in computing $\rho_{\lambda,\mu}$ for the first few stars of the reciprocal wave vectors.  The most complex terms are the three-body terms $\rho_{\lambda,\mu,\nu}$.  In principle, the computation of $\rho_{\lambda,\mu,\nu}$ is also straightforward with Monte Carlo methods, but in this case problems might arise in terms of the accuracy of such three-body correlation functions and the number of terms necessary in order to get convergence for the bound on $f_{s}$.  Only an explicit computation will be able to answer this.  Often one avoids the explicit computation of three-body correlations by a suitable approximation in terms of correlation functions of lower order -- for instance, the so-called convolution approximation.\cite{Feenberg}.  In conclusion, with present computational resources it should be possible to perform the computations needed to get an improved bound on $f_{s}$.  

\section{CONCLUSIONS}

Ref.~\onlinecite{GalliRS} calculated an upper bound for the superfluid fraction using an optimized one-body phase function and state-of-the-art Monte Carlo one-body densities, finding an upper bound for the superfluid fraction of 20\% for solid $^4$He near the melting pressure, 
significantly higher than has been observed.  Given an additional degree of freedom by permitting a many-body phase function, it should be possible to lower the flow energy for the case of the solid, and therefore to lower the corresponding superfluid fraction, perhaps by a considerable amount.  
The present work develops the theory for the superfluid fraction in a periodic solid including a two-body phase function.  The results should also be applicable to 
condensates in optical lattices, and the 
approach should also apply to disordered solids.
We have shown which quantities are needed for such a computation, which should give information on backflow effects in the solid phase.  Computation of the necessary inputs -- the fourier components of two- and three-body correlation functions -- appears feasible with present computational resources, but this is left as a future development.  

\section*{ACKNOWLEDGMENTS}
This work was supported 
by the Department of Energy through grant 
DE-FG02-06ER46278.  We would also like to acknowledge the hospitality of the 
Kavli Institute for Theoretical Physics and the Aspen Institute of Physics.


\begin{thebibliography}{99}
\bibitem{Leggett1} A. J. Leggett, Phys. Rev. Lett. 25, 1543 (1970).  
\bibitem{KCvycor} E. Kim and M. H. W. Chan, Nature 427, 225 (2004). 
\bibitem{KCgold} E. Kim and M. H. W. Chan, J. Low Temp. Phys. 138, 859 (2005). 
\bibitem{KCbulk} E. Kim and M. H. W. Chan, Science 305, 1941 (2004). 
\bibitem{Reppy} A.-S. C. Rittner and J. D. Reppy, cond-mat/0604528.
\bibitem{Shira} M. Kondo, S. Takada, Y. Shibayama, and K. Shirahama, cond-mat/0607032. 
\bibitem{Kubo} M. Kubota, private communications.
\bibitem{GalliRS} D.E. Galli, L. Reatto, and W. M. Saslow, in preparation.
\bibitem{A&L} A. F. Andreev and I. M. Lifshitz, Sov. Phys. JETP 29, 1107 (1969).
\bibitem{chester} G.V. Chester, Phys. Rev. A 2, 256 (1970).
\bibitem{reatto} L. Reatto, Phys. Rev. 183, 334 (1969).
\bibitem{Andreev} A. Andreev, K. Keshishev, L. Mezkov-Deglin and A. Shal'nikov, Sov. Phys. JETP Lett. 9, 306 (1969). 
\bibitem{Greywall} D. S. Greywall, Phys. Rev. B 16, 1291 (1977). 
\bibitem{Beamish} J. Day, T. Herman, and J. Beamish, Phys. Rev. Lett. 95, 035301 (2005).
\bibitem{Meisel} For a review to 1991, see M. W. Meisel, Physica B 178, 121 (1992).
\bibitem{Leggett2} A. J. Leggett, Physica Fennica 8, 125 (1973).
\bibitem{Leggett3} A. J. Leggett, J. Stat. Phys. 93, 927 (1998).
\bibitem{CeperleyBernu} D. M. Ceperley and B. Bernu, Phys. Rev. Lett. 93, 155303 (2004).
\bibitem{ProkoSvis1} N. Prokofiev and B. Svistunov , Phys. Rev. Lett. 94, 155302 (2005)
\bibitem{GalliRR} D. E. Galli, M. Rossi, and L. Reatto, Phys. Rev. B 71, 140506(R) (2005).
\bibitem{Boninsegni} M. Boninsegni, N. Prokofiev, and B. Svistunov, Phys. Rev. Lett. 96, 105301 (2006) 
\bibitem{GalliPRL} D. E. Galli, and L. Reatto, Phys. Rev. Lett. 96, 165301 (2006).
\bibitem{Saslow1} W. M. Saslow, Phys. Rev. Lett. 36, 1151 (1976). 
\bibitem{Saslow2} W. M. Saslow, Phys. Rev. B 71, 092502 (2005). 
\bibitem{Saslow-hcp} W. M. Saslow and S. Jolad, Phys. Rev. B 73, 092505 (2006). 
\bibitem{KroTym} E. Krotscheck and C. J. Tymczak.  See C. J. Tymczak, Ph.D dissertation, {\it Excitations in Liquid $^4$He Films,} Texas A\&M University (1995).
\bibitem{VitiReattoKalos} S. A. Vitiello, L. Reatto, and M. H. Kalos, in {\it Condensed Matter Theories,} edited by V. C. Aguilera-Navarro, Plenum, New York, 1989.
\bibitem{OrtizCeperley} G. Ortiz and D. M. Ceperley, Phys. Rev. Lett. 75, 4642 (1995).
\bibitem{Feenberg} H. W. Jackson and E. Feenberg, Rev. Mod. Phys. 34, 686 (1962).

\end{thebibliography}
\end{document}